# Optimizing RNA yield using deep neural networks coupled to massively parallel screening

Dinghai Zheng[1], Justin Hong[1], Jun Wang[1], Adrien Villain[2,*], Mickaël Costallat[2], Fernando Ulloa Montoya[2], Vikram Agarwal[1,*]

[1]Data and Computational Sciences, mRNA Center of Excellence, Sanofi, Waltham, MA, USA.
[2]Data and Computational Sciences, mRNA Center of Excellence, Sanofi, Marcy L'Etoile, France.

[*]Corresponding authors: Adrien.Villain@sanofi.com and Vikram.Agarwal@sanofi.com

**Abstract**

Messenger RNA (mRNA)-based therapeutics have emerged as a powerful platform for vaccines, protein replacement therapies, and cancer immunotherapy. A critical bottleneck in mRNA development is manufacturing large quantities of RNA economically, as measured by RNA yield emerging from an *in vitro* transcription (IVT) reaction. However, how promoter-adjacent DNA sequences influence RNA yield remains poorly characterized. Here, we present an integrated deep learning framework that leverages massively parallel next-generation sequencing (NGS) assays to measure RNA yield across large sequence spaces.

A library of $10^5$ randomized oligonucleotide sequences was designed to systematically explore sequence diversity within a defined structural context. DNA and RNA abundances were quantified in parallel using Illumina sequencing, enabling high-resolution measurement of sequence-to-yield relationships at scale. Sequences were one-hot encoded and used to train deep learning models, using a convolutional neural network architecture.

The model achieved a Pearson correlation of 0.94 between predicted and experimentally measured RNA yield on a held-out test set, demonstrating strong generalization across diverse sequence contexts. Importantly, the trained model can be deployed in a production environment to score and rank novel RNA sequence designs by predicted IVT yield, enabling cost-effective, pre-experimental prioritization of the most manufacturable candidates.

This framework establishes a scalable, data-driven approach to DNA and RNA sequence optimization, with broad applicability to vaccine antigen design, therapeutic protein delivery, and synthetic biology. By integrating high-throughput experimentation with advanced deep learning modeling, it significantly reduces screening costs and accelerates RNA engineering cycle times.

## 1 Introduction

Messenger RNA (mRNA)-based therapeutics have emerged as a versatile platform for vaccination against infectious disease, genetic engineering, and cancer therapy [1]. Despite rapid progress, optimizing RNA sequence features that enhance translation efficiency, stability, and manufacturability remains a major challenge [2] [3] [4] [5] [6]. Traditional design approaches rely on predefined rules or limited experimental datasets, which often fail to capture the complex biophysical and biochemical determinants underlying RNA function. Massively parallel assays combined with next-generation sequencing (NGS) now enable high resolution, quantitative evaluation of hundreds of thousands of RNA sequences simultaneously, providing powerful datasets to train predictive deep learning models which can be used to guide rational sequence design.

In this paper, we focus on the problem of producing large amounts of RNA from a given DNA input, which would help minimize manufacturing costs associated with large-scale RNA production, as needed for clinical trials. This cell-free process known as IVT is performed thanks to DNA-dependent RNA polymerases, molecular machines repurposed to catalyze IVT. Typically, single-subunit polymerases derived from bacteriophages (e.g., SP6 or T7) are used due to their ease of production and ability to catalyze high-yield reactions. Sequences downstream of the promoter (the DNA sequence recognized by the enzyme to initiate transcription) have been shown to influence yield and byproduct formation during IVT [7]. We hypothesize that some sequence characteristics (such as GC content, dinucleotide frequency, or other motifs) influence RNA yield. We developed an integrated deep learning framework that leverages massively parallel NGS assay data to guide superior sequence design. Our pipeline includes the development of: (i) a parallelized IVT assay applied to a diverse pool of DNA sequences, (ii) quantitative NGS-based measurement of DNA and RNA abundance, (iii) bioinformatics workflows for data processing, and (iv) deep learning sequence-to-function models trained using these processed measurements. The models learn a function mapping a DNA sequence to its corresponding predicted RNA yield. Model-guided sequence optimization then identifies designs predicted to maximize RNA yield. The system allows design iteration and wet lab validation cycles in an active learning loop, enabling iterative improvement of RNA yield.

# 2 Data and Methods

## 2.1 NGS-Based Experimental Data Generation

A library of $10^5$ random, 200nt-long oligos were designed to include a promoter and a randomized variable region (**Figure 1**). DNA sequencing was performed on two replicates using the KAPA HyperPrep Kit [8] and sequenced with Illumina sequencing (paired-end 2*150nt). In parallel, IVT was performed in a single reaction on the pooled oligo library to obtain RNA. Six RNA replicates were sequenced using SMARTer smRNA-Seq Kit for Illumina [9] following the manufactures' protocols, and sequenced with Illumina sequencing (paired-end, 2*150nt).

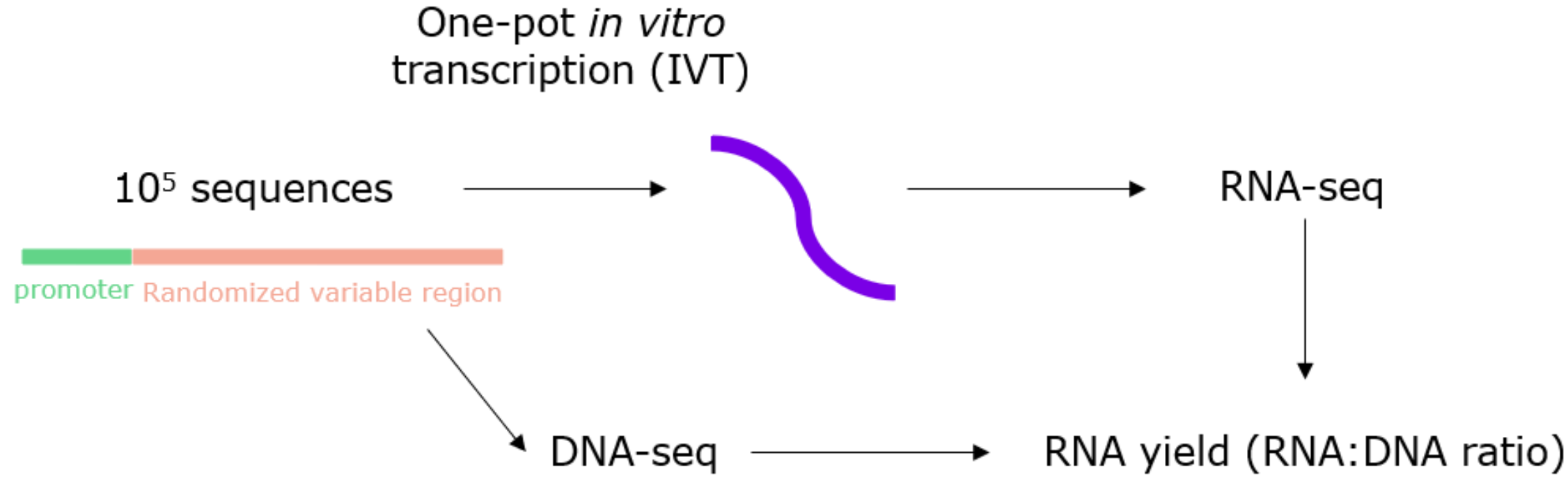

**Figure 1**: Diagram of the NGS-based experimental data generation

### 2.2 NGS data analysis and Machine Learning modeling

DNA and RNA sequencing data were each mapped back to the reference sequences. From the two DNA-seq libraries, 73,754 oligos were detected in total, which had an average read counts of 178.3 and 208.5 reads per oligo, respectively. From the six RNA-seq libraries, reads from 65,707 oligos were detected in total, which had an average read counts of 30.41, 32.5, 52.3, 55.38, 49.55, and 57.07, respectively. We required at least 100 mean DNA read counts across the two DNA libraries to keep oligos, which led to retaining 24,245 oligos in total. Full-length DNA and RNA fragments were quantified, and the $\log_2$(RNA CPM / DNA CPM) ratio, where CPM reflects read depth in counts per million, was calculated to signify the RNA yield emerging from each DNA template.

Deep learning modeling was performed using the convolutional neural network architecture shown (**Figure 2**). The $10^5$ oligos were split into 10 non-overlapping folds. Each fold was used as a test set to train 10 models. Sequences were one-hot encoded and the input sequence length was kept constant. The model architecture follows a simplified version of what is described in prior work [3]. The model contains four convolutional layers (filter size: 5) with three max pooling steps (step size: 2) interleaved with layer norms and an MLP prediction head that maps from convolutional filter outputs to a scalar prediction. Although it is possible to perform a deeper hyperparameter search to optimize the CNN architecture or test other types of architectures altogether, we empirically found that this simple architecture performed well and was fast during training and inference.

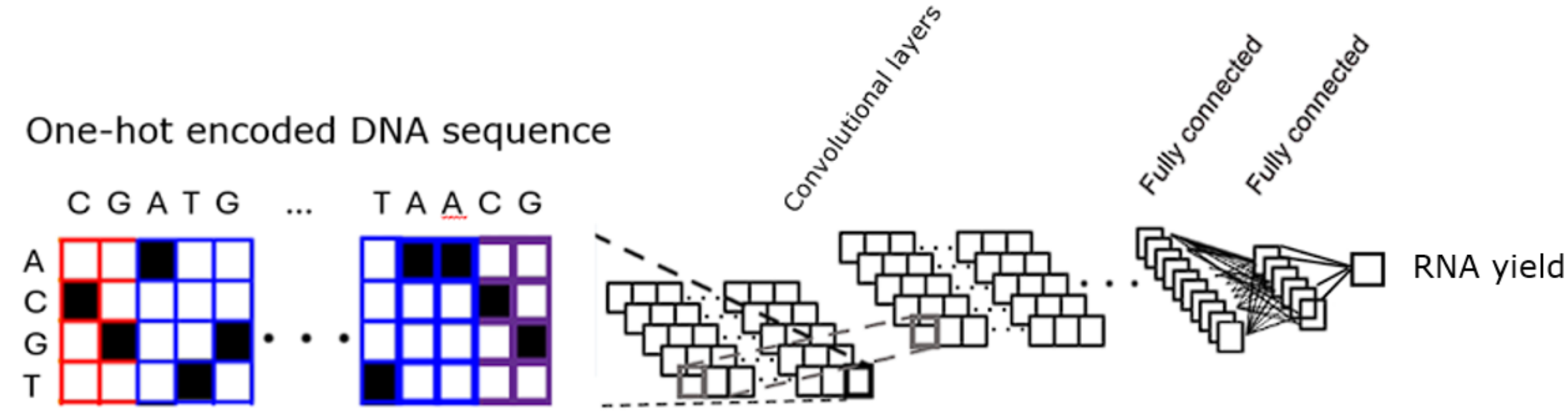

**Figure 2**: Diagram of deep learning architecture used to predict RNA yield.

## 3 Results

### 3.1 Enhanced Sequence Representations and Predictive Accuracy

The trained deep learning models demonstrated strong predictive performance on the held-out test sets, with predicted RNA yield showing a high correlation to experimentally measured RNA yield (R = 0.94, **Figure 3**). This result consolidates data from all 10 folds before computing, although we provide statistics across folds too (Table 1). This result confirms that the models successfully learned meaningful sequence-to-function relationships from the massively parallel NGS dataset, capturing the underlying determinants of RNA yield across a highly diverse sequence space. The high correlation observed across the full range of RNA yield values — from low- to high-yield sequences — indicates that the models generalize well beyond the training data and are not limited to interpolating within a narrow performance range.

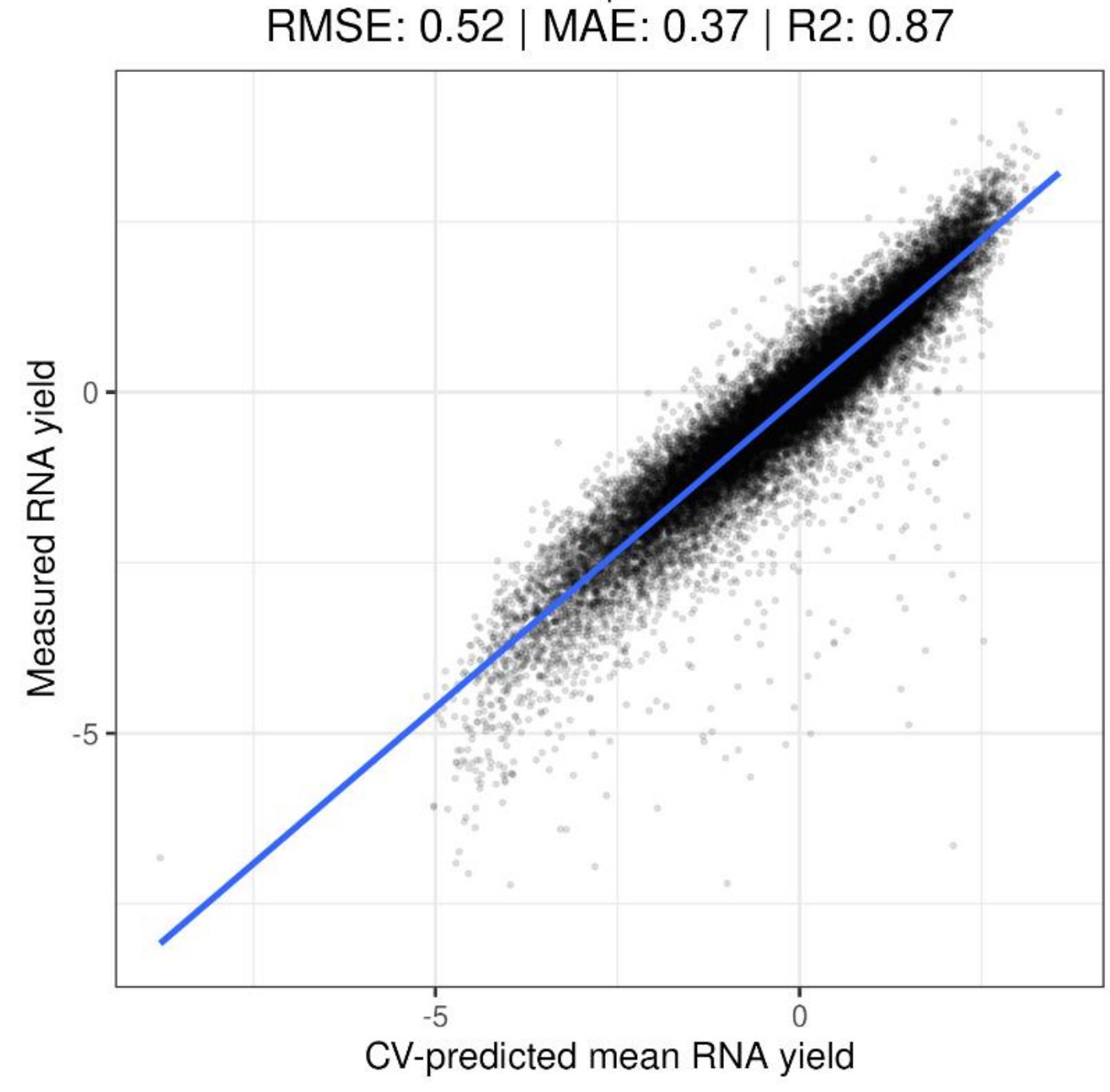

**Figure 3**: Scatter plot evaluating the relationship between measured and predicted RNA yield.

| Statistic | Value |
|---|---|
| PCC_mean | 0.939 |
| PCC_sd | 0.00392 |
| SCC_mean | 0.945 |
| SCC_sd | 0.00338 |
| RMSE_mean | 0.517 |
| RMSE_sd | 0.0185 |
| MAE_mean | 0.372 |
| MAE_sd | 0.00696 |
| R2_mean | 0.873 |
| R2_sd | 0.00945 |

*Table 1: Statistics across all folds; PCC: Pearson Correlation Coefficient, SCC: Spearman's rank Correlation Coefficient; RMSE: Root Mean Squared Error; MAE: Mean Absolute Error; R2: R-squared*

These results support our hypothesis that sequence-encoded features, such as GC content, dinucleotide frequencies, and other local sequence motifs, are informative predictors of RNA yield. By encoding sequences as one-hot representations and training on quantitative NGS-derived abundance ratios, the model was able to extract non-trivial patterns that would be difficult to identify through manual inspection or rule-based approaches.

### 3.2 Model Deployment for RNA Sequence Scoring and Ranking

Beyond its predictive accuracy, the model was designed for practical deployment in a production RNA design environment (**Figure 4**). Given a pool of candidate DNA sequences, the model assigns each sequence a predicted RNA yield score, enabling rapid, cost-free prioritization prior to any wet lab synthesis. Sequences are ranked in decreasing order of predicted IVT yield, allowing design teams to focus experimental resources on the most manufacturable candidates.

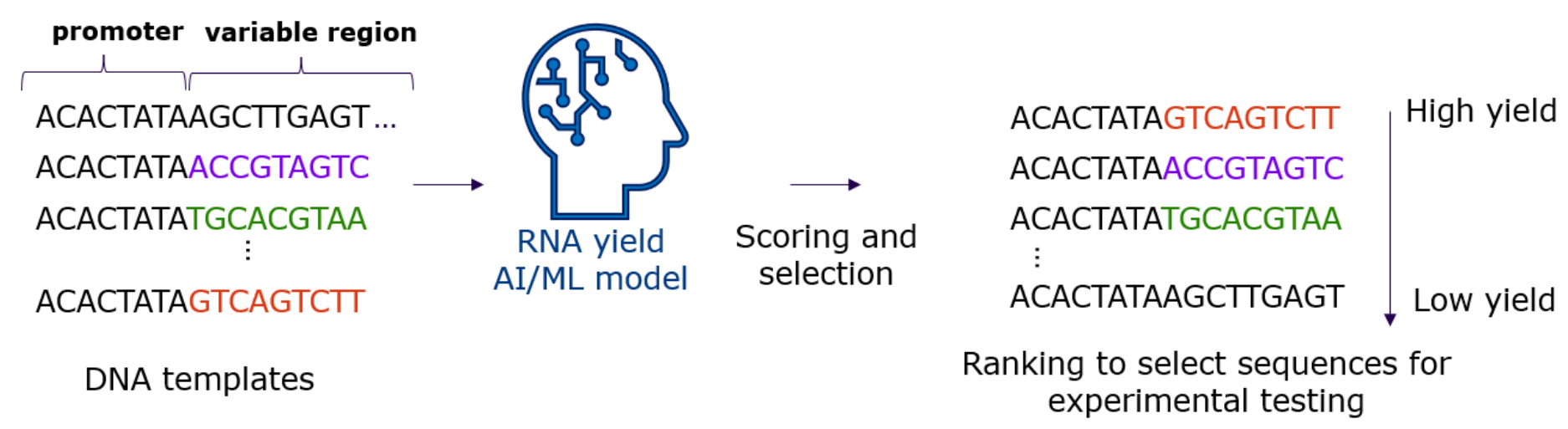

**Figure 4**: Diagram of how the RNA yield model is used in production to score and rank new DNA sequences

This scoring and ranking workflow integrates seamlessly into iterative RNA design cycles: newly proposed sequences—whether generated by combinatorial design or generative AI models—can be filtered through the RNA yield model as an early-stage manufacturability gate. Sequences predicted to fall below a defined yield threshold can be deprioritized or redesigned before committing to synthesis, significantly reducing experimental burden and accelerating the overall engineering timeline.

## 4 Conclusion

We demonstrated how an important manufacturability parameter, RNA yield, can be modeled by deep learning with the support of massively parallel screening assays. The innovation lies in combining massively parallel experimental measurements with advanced deep learning modeling to uncover non-trivial determinants of RNA production. Unlike conventional rule-based or manually curated design strategies, this approach automatically learns predictive patterns from large experimental datasets, capturing subtle sequence-to-function relationships. In order to better understand what the model has learned, a possible extension to this work is to perform positional analysis coupled to motif reconstruction similar to earlier work [3]. Relative to similar work such as [5], our experimental design has the distinct advantages to use completely random oligo sequences, as opposed to naturally occurring sequences, and to use N1-methylpseudouridines to better match the nucleotide composition of mRNA therapeutics.

The framework is adaptable to different functional outputs, making it suitable for optimizing expression kinetics, structural integrity, UTR configurations, and codon context effects. We note that the experimental strategy may suffer certain limitations, such as testing the sequence pool in artificial conditions in which diverse templates are under competition for polymerase, and RNA-RNA interactions may lead to altered polymerase dynamics.

Nevertheless, applications span vaccine antigen expression, therapeutic protein delivery, genetic engineering of cells, and synthetic biology constructs requiring fine-tuned RNA control. Economically, this strategy reduces screening costs while significantly accelerating the RNA engineering cycle times.

### Conflict of interests

The authors are Sanofi employees and may hold shares and/or stock options in the company.

### Funding

This work was funded by Sanofi.